\documentclass{Interspeech}

\interspeechcameraready

\usepackage{amsmath}
\usepackage{amssymb}
\usepackage{cleveref}
\usepackage{enumitem}
\usepackage[left=1.8cm,right=1.8cm,top=3.6cm,bottom=0.8cm]{geometry}
\usepackage{graphicx}
\usepackage{pifont}
\usepackage{tabularx,booktabs}
\usepackage{url}
\usepackage{xcolor}

\definecolor{xiaomi_gray}{HTML}{A9A9A9}
\definecolor{catpuccin_red}{HTML}{f38ba8}
\definecolor{catpuccin_blue}{HTML}{89b4fa}
\newcommand{\cmark}{\ding{51}}%
\newcommand{\code}[1]{\texttt{#1}}

\setlist[enumerate]{leftmargin=2em}

\title{X-ARES: A Comprehensive Framework for Assessing Audio Encoder Performance}

\author[affiliation={1}]{Junbo}{Zhang}
\author[affiliation={1}]{Heinrich}{Dinkel}
\author[affiliation={1}]{Yadong}{Niu}
\author[affiliation={1}]{Chenyu}{Liu}
\author[affiliation={1}]{Si}{Cheng}
\author[affiliation={2}]{Anbei}{Zhao}
\author[affiliation={1}]{Jian}{Luan}

\affiliation{}{MiLM Plus, Xiaomi Inc.}{China}
\affiliation{}{Amazon.com, Inc.}{China}
\email{\{zhangjunbo1, dinkelheinrich, niuyadong\}@xiaomi.com}


\begin{document}

\maketitle

\begin{abstract}
    We introduces X-ARES (eXtensive Audio Representation and Evaluation Suite), a novel open-source benchmark designed to systematically assess audio encoder performance across diverse domains.
    By encompassing tasks spanning speech, environmental sounds, and music, X-ARES provides two evaluation approaches for evaluating audio representations: linear fine-tuning and unparameterized evaluation.
    The framework includes 22 distinct tasks that cover essential aspects of audio processing, from speech recognition and emotion detection to sound event classification and music genre identification.
    Our extensive evaluation of state-of-the-art audio encoders reveals significant performance variations across different tasks and domains, highlighting the complexity of general audio representation learning.
\end{abstract}

\section{Introduction}

The field of audio representation learning has witnessed remarkable progress in recent years~\cite{dinkel2024dasheng,baevski2020wav2vec,baevski2022data2vec}, driven by the increasing availability of audio data and advancements in deep learning methodologies.
Effective audio encoders, capable of transforming raw audio waveforms into meaningful representations, are crucial for a wide range of applications, including speech recognition, environmental sound analysis, music information retrieval, and multimodal approaches combining audio and large language models (LLMs) ~\cite{chu2023qwen}.
While recent research has explored discrete audio representations and tokenization methods~\cite{kyutai2024moshi}, there remains a notable gap in the availability of general audio embeddings that can effectively serve a broad range of downstream tasks~\cite{wang2024comparative}.

While benchmarks like HEAR~\cite{turian2022hear}, SUPERB~\cite{yang2021superb}, and DASB~\cite{mousavi2024dasb} have contributed to the evaluation of audio models, there remains a need for benchmarks that comprehensively assess encoder capabilities across a wider range of tasks and evaluation paradigms, particularly focusing on real-world applicability.

To address these limitations, we introduce X-ARES (eXtensive Audio Representation and Evaluation Suite), a novel open-source benchmark designed for the rigorous evaluation of audio encoder capabilities.
X-ARES aims to provide a comprehensive and standardized platform for assessing and comparing audio encoders, facilitating advancements in audio representation learning and promoting the development of robust and versatile models for real-world applications.
The key contributions of this work are as follows:

\begin{enumerate}
    \item We present X-ARES, a comprehensive benchmark suite that evaluates audio encoders across a diverse set of tasks spanning speech, environmental sounds, and music domains.
    \item We introduce two complementary evaluation methodologies: parameterized multilayer perceptron (MLP) and unparameterized k-nearest neighbors (k-NN), providing a more nuanced assessment of encoder performance.
    \item We provide an extensive evaluation of state-of-the-art audio encoders using X-ARES, showing their relative strengths and weaknesses.
    \item We release X-ARES as an open-source toolkit, facilitating easy integration of new encoders and tasks, and promoting reproducibility in audio representation research.
\end{enumerate}

\section{Related Work}

\begin{figure*}[tb]
    \centering
    \includegraphics[width=0.8\linewidth]{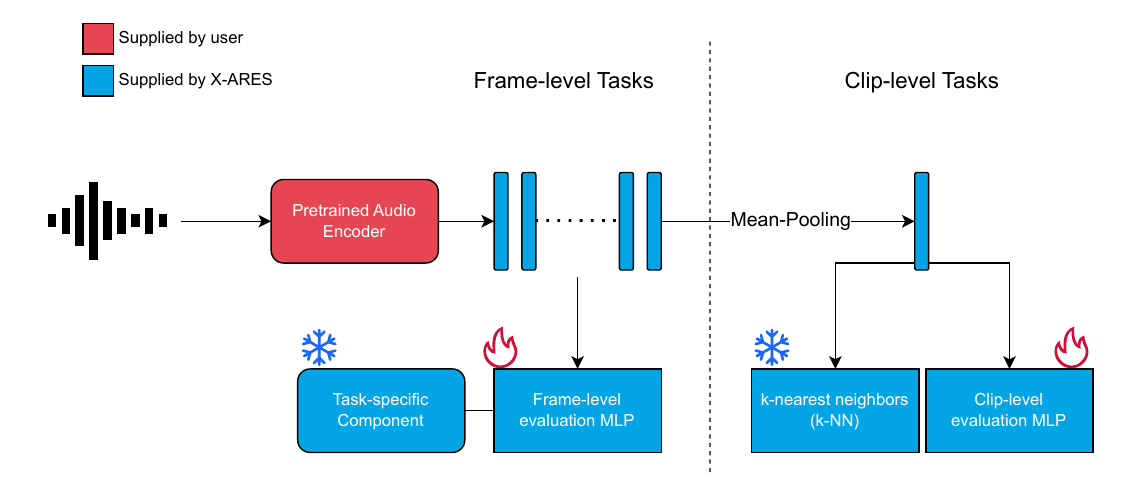}
    \caption{The proposed X-ARES framework. Users provide a single pretrained audioencoder, which outputs frame-level embeddings. Embeddings are evaluated using a fine-tuned MLP layer for clip- and frame-level tasks. Further a non-parameterized kNN algorithm is used to evaluate the quality of embeddings. For specialized tasks, pre-trained decoders are incorporated as task-specific components.}
    \label{fig:xares}
\end{figure*}

\subsection{HEAR: Holistic Evaluation of Audio Representations}
X-ARES is strongly inspired by the HEAR benchmark~\cite{turian2022hear}, which assesses audio representations across environmental sound and music tasks.
While HEAR provides an excellent foundation, X-ARES introduces several enhancements:

\vspace{-10pt}
\paragraph*{Unified performance evaluation} Performance for frame- and clip-level tasks in HEAR is evaluated using different model heads, effectively creating two distinct modes: one for fine-grained, frame-level analysis and another for coarse, clip-level analysis. As a result, solutions vary significantly between evaluation schemes, making results across tasks inherently incomparable.
X-ARES addresses this issue by streamlining the pipeline, requiring users to provide a single embedding sequence.

\vspace{-10pt}
\paragraph*{Focus on real-world applications}  HEAR comprises 19 tasks in total, 17 of which are unique, while two tasks differ only in their available training data.
While the tasks in HEAR encompass various application scenarios for sound event detection and music processing, they lack variety in human voice processing.
X-ARES offers a more comprehensive and balanced distribution of tasks across human voice, music, and environmental sound domains,
leveraging a suite of open-source datasets that reflect real-world scenarios and user experiences.
Further, some tasks in HEAR, may have limited applications and high variance during testing (e.g., Gunshot Triangulation and Beehive) due to the factors such as small sample sizes, which have led to many follow-up works discarding those tasks.

\vspace{-10pt}
\paragraph*{Dual evaluation methods}
In addition to linear projection like HEAR, we also utilize unparameterized methods for classification.
This evaluation aims at investigating the use of features for cases such as unsupervised clustering.

\vspace{-10pt}
\paragraph*{More efficient system} X-ARES implements several key optimizations for improved evaluation efficiency.
We utilize WebDataset~\cite{webdataset} for data loading, achieving 3-5x faster processing speeds compared to traditional approaches, being effective even on low-cost hard disk drives.
All datasets are pre-packaged in standardized tar format on Zenodo, ensuring reproducibility and simplified preparation.
The framework provides a unified embedding interface where users only need to provide a single frame-level embedding sequence.
Our TaskConfig system enables flexible configuration of evaluation parameters without code modifications.

\subsection{SUPERB: Speech processing Universal PERformance Benchmark}
SUPERB~\cite{yang2021superb} and its derivatives primarily focus on speech processing tasks using self-supervised learning (SSL) representations.
In recent years, SUPERB also included additional tasks such as emotion recognition and sound codecs, but notably, it does not include environmental audio or music related tasks.
X-ARES broadens this scope with the inclusion of non-speech related tasks (environmental audio, music), enabling a more comprehensive evaluation of audio representations.

\subsection{DASB: Discrete Audio and Speech Benchmark}
DASB~\cite{mousavi2024dasb} benchmarks discrete audio tokens across various tasks, mainly focuses on the speech domain.
While discretization is an important research field, continuous representations offer complementary advantages.
Continuous representations directly addresses the need for robust audio encoders in multimodal applications, where continuous embeddings are often preferred for seamless integration and efficient processing~\cite{wang2024comparative,yu2024salmonn}.
The output of X-ARES can be used to complement discrete representation research by, for example, injecting general semantic information into codecs~\cite{kyutai2024moshi},
or evaluating the loss of information during the discretisation process.

\section{Framework Design}

X-ARES~\footnote{\url{https://github.com/jimbozhang/xares}} is designed as a flexible and extensible framework for evaluating audio encoders across a diverse range of tasks.

\begin{table*}[htb]
    \centering
    \caption{Overview of tasks in X-ARES benchmark. All provided tasks use a MLP as fine-tuning method, while a subset also supports k-NN evaluation. Tasks denoted with $^\clubsuit$ use a stratified training subset. For all metrics, higher is better.}
    \label{tab:testset}
    \begin{tabularx}{\textwidth}{Xlllrcc}
        \toprule
        \textbf{Domain} & \textbf{Dataset}                                        & \textbf{Task Type}          & \textbf{Metric} & \textbf{\#} & \textbf{Frame-level} & \textbf{k-NN} \\
        \midrule
        \textbf{Speech} & ASV2015~\cite{kinnunen2018automatic}                    & Spoofing detection          & Acc             & 2           &                      & \cmark        \\
                        & CREMA-D~\cite{cao2014crema}                             & Emotion recognition         & Acc             & 5           &                      & \cmark        \\
                        & Fluent Speech Commands~\cite{lugosch2019speech}         & Intent classification       & Acc             & 248         &                      & \cmark        \\
                        & LibriCount~\cite{stoter2018libricount}                  & Speaker counting            & Acc             & 11          &                      & \cmark        \\
                        & LibriSpeech-100h~\cite{p2015ls}                         & Speech recognition          & iWER            & -           & \cmark               &               \\
                        & LibriSpeech-MF~\cite{p2015ls}                           & Gender classification       & Acc             & 2           &                      & \cmark        \\
                        & RAVDESS~\cite{livingstone2018ryerson}                   & Emotion recognition         & Acc             & 8           &                      & \cmark        \\
                        & VocalSound~\cite{gong_vocalsound}                       & Non-speech sounds           & Acc             & 6           &                      & \cmark        \\
                        & Speech Commands V1~\cite{warden2018speech}              & Keyword spotting            & Acc             & 30          &                      & \cmark        \\
                        & VoxCeleb1~\cite{nagrani2020voxceleb}                    & Speaker identification      & Acc             & 1251        &                      & \cmark        \\
                        & VoxLingua33$^\clubsuit$~\cite{valk2021voxlingua107}     & Language identification     & Acc             & 33          &                      & \cmark        \\
        \midrule
        \textbf{Sound}  & Clotho~\cite{drossos2020clotho}                         & Sound retrieval             & Recall@1        & -           &                      &               \\
                        & DESED~\cite{turpault2019sound}                          & Sound event detection       & Segment-F1      & 10          & \cmark               &               \\
                        & ESC-50~\cite{piczak2015esc}                             & Environment classification  & Acc             & 50          &                      & \cmark        \\
                        & FSD18-Kaggle~\cite{fonseca2018general}                  & Sound event detection       & mAP             & 41          &                      &               \\
                        & FSD50k~\cite{fonseca2021fsd50k}                         & Sound event detection       & mAP             & 200         &                      &               \\
                        & UrbanSound 8k~\cite{salamon2014dataset}                 & Urban sound classification  & Acc             & 10          &                      & \cmark        \\
                        & Vocal Imitation~\cite{kim2018vocal}                     & Onomatopoeia classification & Acc             & 302         &                      & \cmark        \\
        \midrule
        \textbf{Music}  & Free Music Archive (FMA) Small~\cite{defferrard2016fma} & Music genre classification  & Acc             & 8           &                      & \cmark        \\
                        & GTZAN Genre~\cite{sturm2013gtzan}                       & Genre classification        & Acc             & 10          &                      & \cmark        \\
                        & MAESTRO$^\clubsuit$ ~\cite{hawthorne2018enabling}       & Note classification         & Segment-F1      & 88          & \cmark               &               \\
                        & NSynth-Instruments$^\clubsuit$ ~\cite{nsynth2017}       & Instruments Classification  & Acc             & 11          &                      & \cmark        \\
        \bottomrule
    \end{tabularx}
\end{table*}

\subsection{Overall Architecture}
The X-ARES framework, illustrated in \Cref{fig:xares}, offers an automated pipeline to comprehensively evaluate pretrained audio encoders.
X-ARES employs two distinct evaluation methodologies: MLP (Linear Fine-Tuning) and k-NN (Unparameterized Evaluation).
For MLP evaluation, the user-provided encoder and Task-Specific Components are frozen, and only a linear MLP is trained to assess the representation quality.
In contrast, k-NN evaluation is fully unparameterized, classifying extracted embeddings based on their proximity in the feature space.

\subsection{Task Configuration and Data Processing}

X-ARES uses a flexible \texttt{TaskConfig} system to define evaluation tasks,
and leverages WebDataset as a high-performance data loading framework, offering significant advantages in handling large-scale audio datasets, particularly on mechanical hard drives.
WebDataset uses tar archives to enable efficient, sequential data access with minimal seek operations.
We have meticulously packaged all datasets in standard tar format and uploaded them to Zenodo, creating a universally accessible resource that extends beyond X-ARES's immediate use.

\subsection{Task-Specific Components}

For some specialized tasks, X-ARES provides pre-trained models as fixed components.
For the audio captioning tasks, we utilize the \texttt{google-bert/bert-base-uncased}~\cite{turc2019} model from Hugging Face as a pre-trained text encoder.
For the speech recognition tasks, we employ the \texttt{Qwen/Qwen2.5-0.5B}~\cite{yang2024qwen2} model as a decoder to generate text from audio representations.
These pre-trained models are used with frozen parameters. During the training process, only the parameters of the MLP adapter layers are updated.

\subsection{User-Provided Audio Encoder}
X-ARES requires users to provide their audio encoder, which should be implemented as a standard \code{torch.nn.Module}.

Frame-level embeddings are extracted from an audio sample with batch shape $\mathcal{R}^{B \times W}$, where $B$ denotes the batch size and $W$ the number of audio samples.
The user defined encoder should output a tensor of shape $\mathcal{R}^{B \times T \times D}$, where $T$ represents the number of encoded features and $D$ is the embedding dimension.
Users further need to provide the resolution of $T$, given in milliseconds.

To aid users in verifying the compliance of their encoders, X-ARES provides a dedicated checking utility.
Furthermore, to facilitate the integration process and offer practical guidance, X-ARES includes example wrappers demonstrating how to encapsulate common open-source audio encoders to meet the framework's requirements.

\section{Task Categories}

An overview of X-ARES tasks across three fundamental audio domains: speech, environmental sounds, and music can be seen in \Cref{tab:testset}.

Speech tasks in X-ARES assess both linguistic content (e.g., speech content, word spotting) and paralinguistic features (e.g., emotion, speaker identity, accent).
Tasks like speech recognition (Librispeech-100h~\cite{p2015ls}), speaker identification (VoxCeleb1~\cite{nagrani2020voxceleb}), language identificaton (VoxLingua107~\cite{valk2021voxlingua107}), synthesized speech detection (ASV2015~\cite{kinnunen2018automatic}) and emotion recognition (CREMA-D~\cite{cao2014crema}, RAVDESS~\cite{livingstone2018ryerson}) are included to evaluate the encoder's ability to capture fine-grained paralinguistic features, which are crucial for applications in voice assistants and affective computing.

Environmental sound tasks evaluate acoustic event detection and scene classification capabilities in diverse real-world settings, from urban environments to vehicle acoustics.
X-ARES includes tasks such as environment classification (ESC-50~\cite{piczak2015esc}, Urbansound8k~\cite{salamon2014dataset}) and sound event detection (FSD50k~\cite{fonseca2021fsd50k}, FSD18-Kaggle~\cite{fonseca2018general}), which are essential for applications in smart cities and environmental monitoring.
Additionally, Clotho~\cite{drossos2020clotho} is included to evaluate the encoder's ability to perform contrastive learning, which is crucial for applications in audio search and recommendation systems.

Music tasks focus on both high-level attributes (genre, mood) and structural elements (tempo, key, beat), covering the essential aspects of music understanding.
X-ARES includes tasks such as genre classification (GTZAN Genre~\cite{sturm2013gtzan}, Free Music Archive~\cite{defferrard2016fma}), instrument classification (NSynth-Instruments~\cite{nsynth2017}) and note classifation(MAESTRO~\cite{hawthorne2018enabling}) which are crucial for music information retrieval and recommendation systems.

The training sets for three tasks — MAESTRO, Nsynth-Instrument, and VoxLingua33 — were sampled in a stratified manner to reduce data size and minimize training time.
For instance, VoxLingua33 was sampled from VoxLingua107, selecting only 33 out of the 107 available languages, as the test set in VoxLingua provides labels only for these 33 languages.

\begin{figure}[tb]
    \centering
    \includegraphics[width=1.1\linewidth]{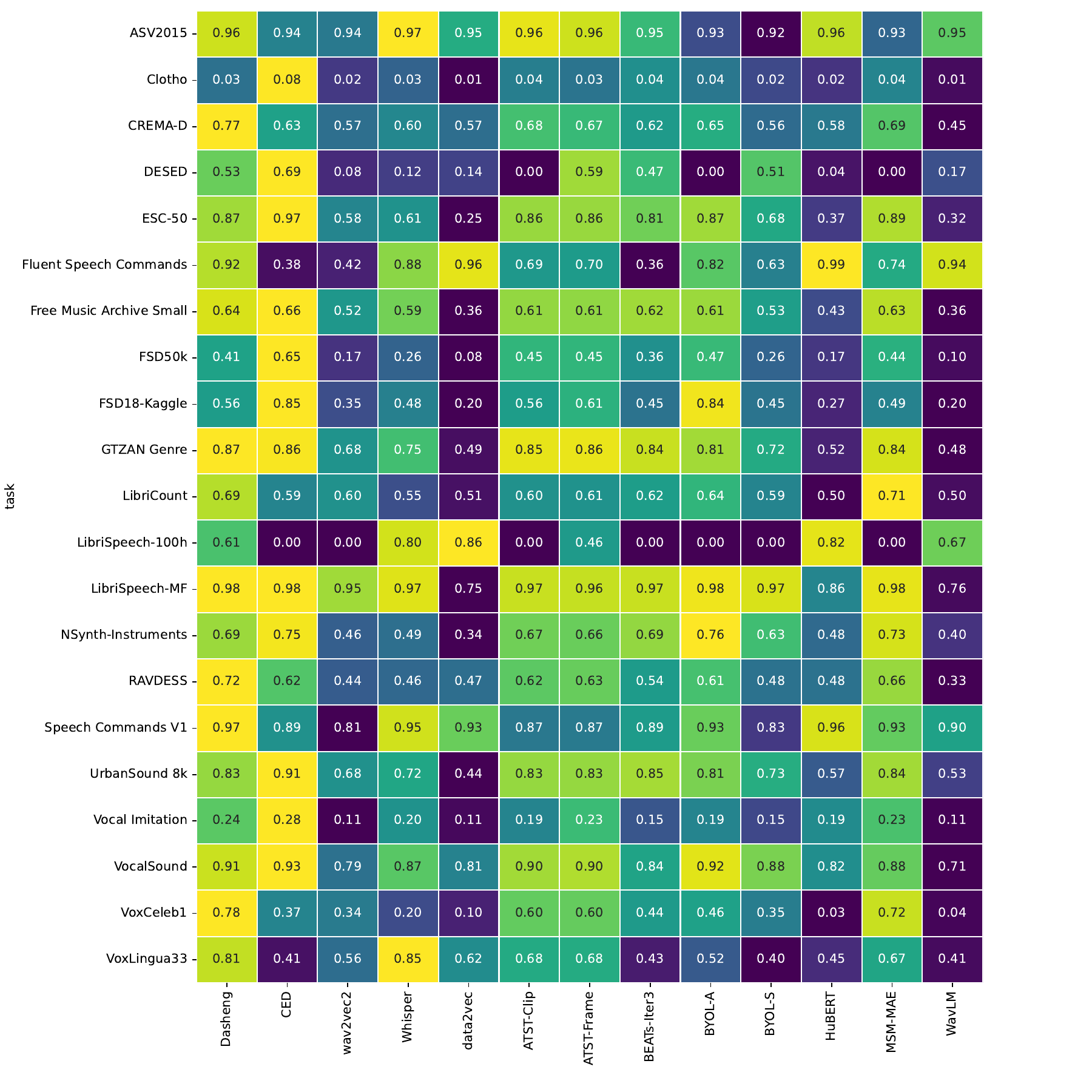}
    \caption{MLP evaluation results for each model and task, where higher is better. }
    \label{fig:mlp_results}
\end{figure}

\begin{figure}[tb]
    \centering
    \includegraphics[width=1.1\linewidth]{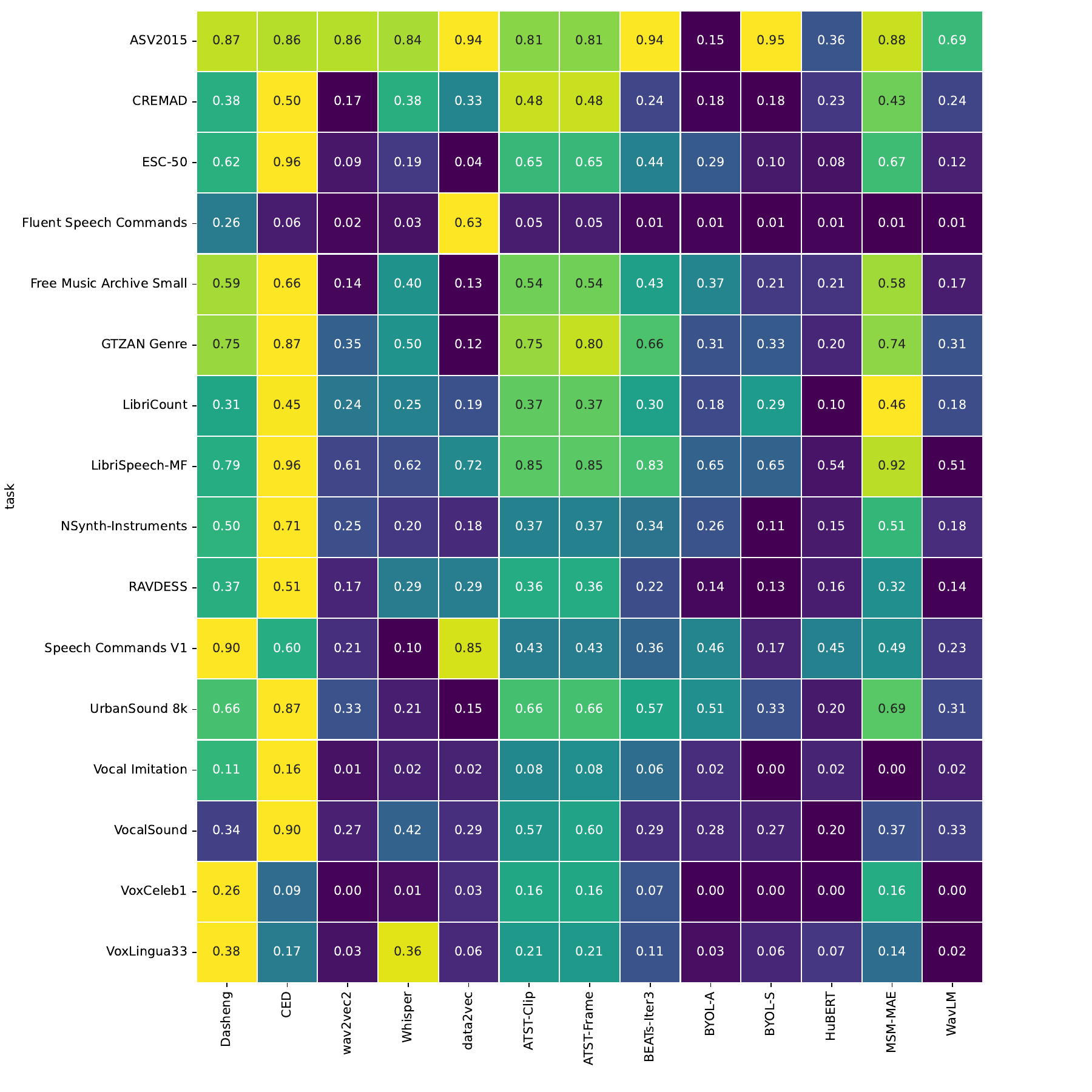}
    \caption{k-NN evaluation results for each model and task, where higher is better.}
    \label{fig:knn_results}
\end{figure}

\section{Evaluation Metrics}

\subsection{Task-Specific Metrics}

A summary of all metrics used in X-ARES is provided in \Cref{tab:testset}.
Accuracy (Acc) is used for multi-class classification tasks across all domains, while Mean Average Precision (mAP) is used for multi-class multi-label classification.
Segment-F1 serves as a metric to assess frame-level performance on a coarse scale, where F1-scores are computed for one-second segments.
For speech recognition tasks, we use an inverted word error rate (iWER), defined as $\text{iWER} = \max(1 - \text{WER}, 0)$, to ensure that higher values correspond to better performance, which is true for all metrics.
Recall@1 is a specialized metric for sound-event retrieval, with Recall@1 representing the average top-1 retrieval performance for both audio-to-text and text-to-audio tasks.

\subsection{Metric Normalization}

To enable comparison across different tasks and metrics, X-ARES normalizes all task-specific metrics to a 0-1 scale:

\begin{equation}
    \hat{M}_i = \frac{M_i - M_i^{\text{min}}}{M_i^{\text{max}} - M_i^{\text{min}}},
\end{equation}
where $\hat{M}_i$ is the normalized metric for task $T_i$, $M_i$ is the raw metric value, and $M_i^{\text{min}}$ and $M_i^{\text{max}}$ are the worst and best possible values of the metric, respectively.
This normalization ensures that performance on different tasks can be meaningfully compared and aggregated.

To calculate the average performance across all tasks, we use a weighted average score across all results:

\begin{equation}
    S = \frac{\sum_{i=1}^{N_{\text{task}}} n_i \hat{M}_i}{\sum_{i=1}^{N_{\text{task}}} n_i},
\end{equation}
where $S$ is the weighted average score, $N_{\text{task}}$ is the total number of tasks, $n_i$ is the size of the test set for task $T_i$, and $\hat{M}_i$ is the normalized metric for task $T_i$.

\section{Experiments and Results}

We run X-ARES over a plethora of publicly available audio-encoders, which can be categorized into three domains.
First, speech encoders such as data2vec~\cite{baevski2022data2vec}, HuBERT~\cite{hsu2021hubert}, wav2vec2-large~\cite{baevski2020wav2vec}, WavLM~\cite{chen2022wavlm} and Whisper-base~\cite{Whisper}.
Second, sound encoders such as BEATs~\cite{chen2022beats}, BYOL-S~\cite{turian2022hear} and CED-base~\cite{dinkel2024ced}.
Third, general audio encoders, such as ATST-Clip~\cite{li2024self}, ATST-Frame~\cite{li2024self}, BYOL-A~\cite{niizumi2023masked}, Dasheng-base~\cite{dinkel2024dasheng} and MSM-MAE~\cite{niizumi2023masked}.
The averaged results are presented in \Cref{tab:averaged_performance}, whereas in the following we focus on each of the two evaluation frameworks.

\begin{table}[h]
    \centering
    \footnotesize
    \begin{tabular}{l||cc}
        \toprule
        Model                               & MLP            & k-NN           \\
        \midrule
        ATST-Clip~\cite{li2024self}         & 0.481          & 0.433          \\
        ATST-Frame~\cite{li2024self}        & 0.622          & 0.437          \\
        BEATs~\cite{chen2022beats}          & 0.451          & 0.361          \\
        BYOL-A~\cite{niizumi2023masked}     & 0.487          & 0.239          \\
        BYOL-S~\cite{turian2022hear}        & 0.391          & 0.259          \\
        CED~\cite{dinkel2024ced}            & 0.496          & \textbf{0.548} \\
        Dasheng~\cite{dinkel2024dasheng}    & \textbf{0.696} & 0.499          \\
        data2vec~\cite{baevski2022data2vec} & 0.565          & 0.379          \\
        HuBERT~\cite{hsu2021hubert}         & 0.602          & 0.207          \\
        MSM-MAE~\cite{niizumi2023masked}    & 0.444          & 0.437          \\
        wav2vec2~\cite{baevski2020wav2vec}  & 0.384          & 0.254          \\
        WavLM~\cite{chen2022wavlm}          & 0.525          & 0.437          \\
        Whisper~\cite{Whisper}              & 0.646          & 0.301          \\
        \bottomrule
    \end{tabular}
    \caption{Weighted average performance for each task, regarding MLP and k-NN evaluation.}
    \label{tab:averaged_performance}
\end{table}

\subsection{MLP results}

Analyzing the MLP results in \Cref{fig:mlp_results} reveals significant discrepancies between the utilized audio encoders.
As expected, speech encoders perform well on ASR and related tasks such as keyword spotting. One particularly strong model is Whisper, achieving scores of 0.80 for ASR, 0.85 for language identification (VoxLingua33), and 0.95 for speech commands. However, the performance of all speech encoders drops sharply in sound and music evaluation scenarios.

In contrast, sound event encoders excel across all sound-event and music-related tasks. Notably, CED performs best on VocalSound (0.93), Clotho (0.08), ESC-50 (0.97), and Urbansound8k (0.91).
However, these models struggle with speech-related tasks, with all tested sound-event encoders scoring 0.0 on the ASR benchmark.

General audio encoders balance the strengths and weaknesses of specialized models. Though they fall short in ASR compared to speech encoders, they offer well-rounded performance, making them ideal for newcomers. Notable models like ATST-Frame and Dasheng perform well across tasks, excelling in speaker identification, emotion recognition, and music genre classification.

\subsection{k-nearest neighbors results}

\Cref{fig:knn_results} present the results of our k-NN evaluations.
Here the results contrast somewhat previous findings.
Performance drops significantly, primarily due to the unparameterized nature of the setting.
Moreover, certain models, such as Wav2Vec2, exhibit poor performance across most tasks.
This suggests that Wav2Vec2 relies heavily on parameterized fine-tuning and does not inherently provide well-balanced features.
A key observation in k-NN evaluation is that sound and general-purpose encoders perform well across the majority of tasks, significantly outperforming speech encoders.
One possible explanation for this behavior is that speech encoders are typically trained for frame-level classification, whereas our k-NN scheme operates at the utterance level.
Additionally, sound-event encoders are exposed to a more diverse range of data, making their features more resilient to variations in input data.

\section{Conclusion}

We presents X-ARES, a comprehensive framework for evaluating audio encoder performance that addresses critical limitations in existing benchmarks.
By designing a diverse task set across speech, environmental sound, and music domains, and implementing both MLP and kNN evaluation methods, we provide a more holistic approach to assessing audio representations.
Our experimental results demonstrate substantial performance differences among state-of-the-art audio encoders.


\bibliographystyle{IEEEtran}
\bibliography{paper}

\end{document}